\documentclass[12pt]{article}
\usepackage{scicite}
\usepackage{times}
\usepackage{graphicx}
\usepackage{mathtools}

\newenvironment{sciabstract}{%
\begin{quote} \bf}
{\end{quote}}

\newcounter{lastnote}

\title{Monitoring Akkuyu Nuclear Reactor Using Antineutrino Flux Measurement} 

\author
{Sertac Ozturk$^{1,2}$, Aytul Adiguzel$^{3}$, V. Erkcan Ozcan$^{3}$ and Gokhan Unel$^{4}$\\
\\
\normalsize{$^{1}$Department of Physics, Gaziosmanpasa University, Tokat, Turkey}\\
\normalsize{$^{2}$Department of Physics and Astronomy, The University of Iowa, Iowa City, IA, USA}\\
\normalsize{$^{3}$Department of Physics, Bogazici University, Istanbul, Turkey}\\
\normalsize{$^{4}$Department of Physics and Astronomy, The University of California, Irvine, CA, USA}\\
\\
\normalsize{E-mail:  sertac.ozturk@gop.edu.tr}
}

\date{\today}

\begin{document} 

\baselineskip24pt
\maketitle

\begin{sciabstract}
  We present a simulation-based study for monitoring Akkuyu Nuclear Power Plant's activity using 
  antineutrino flux originating from the reactor core. A water Cherenkov detector has been designed and 
  optimization studies have been performed using Geant4 simulation toolkit. A first study for the design of a 
  monitoring detector facility for Akkuyu Nuclear Power Plant is discussed in this paper. 
\end{sciabstract}

\section{Introduction}

The first nuclear power plant in Turkey will be constructed at Akkuyu, in Mersin province. 
  Its  operation is planned to start in $2023$. Akkuyu Nuclear Power Plant (NPP) will have 4 power 
  units and each unit will have the capacity of $1.2$ GW. Enriched uranium dioxides ($^{235}U$) 
  will be used as fuel. 

 Measuring antineutrino flux from a nuclear reactor can provide real time information of the status of the reactor and its thermal power. The thermal power produced in the fission process is directly related 
  with emitted antineutrino flux. The relation between neutrino event rate at the detector ($N_{\nu}$) and reactor thermal power ($P_{th}$) can be expressed by $N_{\nu} = \gamma (1+k) P_{th}$ , where $\gamma$ is a constant that depends on the detector (target mass, detection efficiency, etc.) and $k$ is the time dependent factor which takes into account the time evolution of the fuel composition \cite{atom1}.  This property makes a compact antineutrino detector a powerful tool for monitoring a nuclear reactor. 
  
\begin{figure}[!ht]
  \centering
    \includegraphics[width=0.5\textwidth]{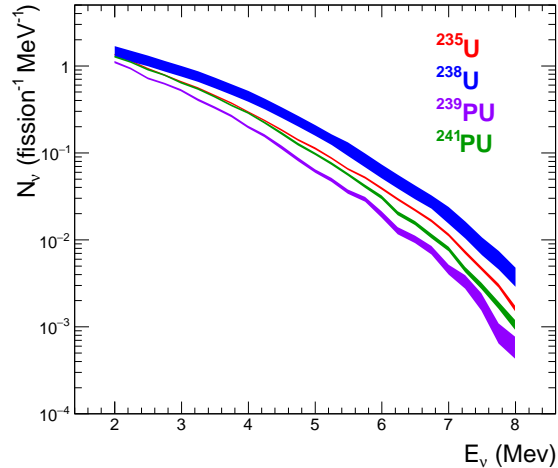}
    \caption{Predicted antineutrino spectra with their relative errors versus neutrino energy. \label{fig-spec}}
\end{figure}

Nuclear reactors are an intense source of antineutrinos. Each fission process releases around 200 MeV energy, 6 $\bar{\nu_e}$ and neutrons. Emitted antineutrino flux by a $1.2$ GW nuclear reactor is about $2.5\times10^{20}$  $\bar{\nu_e}/s$. Predicted emitted neutrino spectra for different nuclear fuels are shown in Figure \ref{fig-spec}. The spectra for $^{235}U$, $^{239}Pu$ and $^{241}Pu$ isotopes have been converted from ILL electron data \cite{ILL}, and the $^{238}U$ antineutrino spectra has been taken from an ab initio calculation \cite{abinit}.
 
Neutrino interaction rate can be estimated approximately using \cite{vinter} :

\begin{equation}
R_\nu = \frac{N_f N_p <\sigma>}{4 \pi L^2}
\end{equation}
where $N_p$ is the number of protons in the target medium, $L$ is the distance between reactor core and detector, 
$N_f $ is the average fission rate given by the equation:

\begin{equation}
N_f = 6.24\times 10^{18} (\frac{P_{th}}{MW}) (\frac{MeV}{W_{e}}) s^{-1}
\end{equation}
where $W_{e}$=203.78 MeV is average energy release per fission, $<\sigma> = 5.8\times10^{-43}$ $cm^{2}$ is the average cross section. 
For a water target and $1.2$ GW reactor thermal power, eq.(1) can be rewritten as

\begin{equation}
R_\nu = 9.86\times 10^{5} (\frac{V}{m^{3}}) (\frac{m^2}{L^2}) events/day
\end{equation}
 
 Considering $L=30$ m and $V=0.96$ $m^3$, we expect around $1050$ events in a day with a detector near Akkuyu NPP.
 
 This figure can be compared to the rough estimation recommended in the IAEA Workshop on Antineutrino Detection for Safeguards Applications report \cite{IAEArepo}, given as: 
 
\begin{equation}
\# events/day = 730 \times MW_{th} \times \frac{V}{L^2} \times \epsilon
\end{equation}

where $\epsilon$ is the detection efficiency. At $\epsilon =1$ the two formulas give very close answers: 1050 from the former and 934 from the latter.
 
\section{Detector Design}

An antineutrino can be detected by charged-current antineutrino-proton scattering, also known as inverse beta decay (IBD):
$\bar{\nu}+p \to e^+ + n$  .
The positron generates the prompt signal, and subsequently the thermal neutron capture process will give a second delayed signal. This delayed  coincidence of the two signals in the time window of 20-80 $\mu s$  is commonly used as trigger for antineutrino detection. 

\begin{figure}[!ht]
  \centering
    \includegraphics[width=0.5\textwidth]{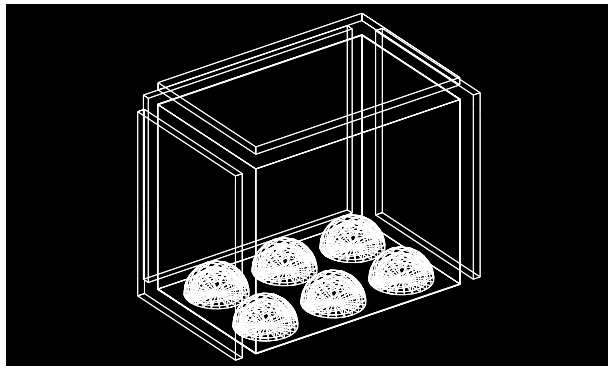}
    \caption{A monitoring unit proposed for Akkuyu nuclear power plant. The hemispheres represent the 10 inch photomultipliers, the rectangles stand for veto scintillators. \label{unit}}
\end{figure}

For the monitoring of Akkuyu Nuclear Reactor, we propose a relatively cheap neutrino detector composed of compact and transportable units. Each unit is planned to be composed of a Gadolinium-doped water Cherenkov detector that can be used for antineutrino detection. A schematic view of such a unit is shown in Figure \ref{unit}. Each detector unit should be divided into two physical regions. Inner region of the detector is planned to be of cubic form with dimensions  $80\times100\times120$ cm, containing about 1 ton of Gadolinium doped water. Gadolinium (Gd) has the highest thermal neutron capture cross section, very suitable for such a detector.  Therefore the Gd-doped water will play the role of the target for the charged-current antineutrino-proton scattering. The outer region of the detector unit could be covered by several layers. The first layer is made of about 3 cm thick plastic scintillator panels to veto cosmic charged particle. The following layers are planned as a passive shielding to suppress neutron and cosmic background. The design and material decision for passive shielding is under study. A large number of such units could be assembled and operated to increase the detection, thus monitoring, efficiency of the overall system. 

\begin{figure}[!ht]
  \centering
    \includegraphics[width=0.47\textwidth]{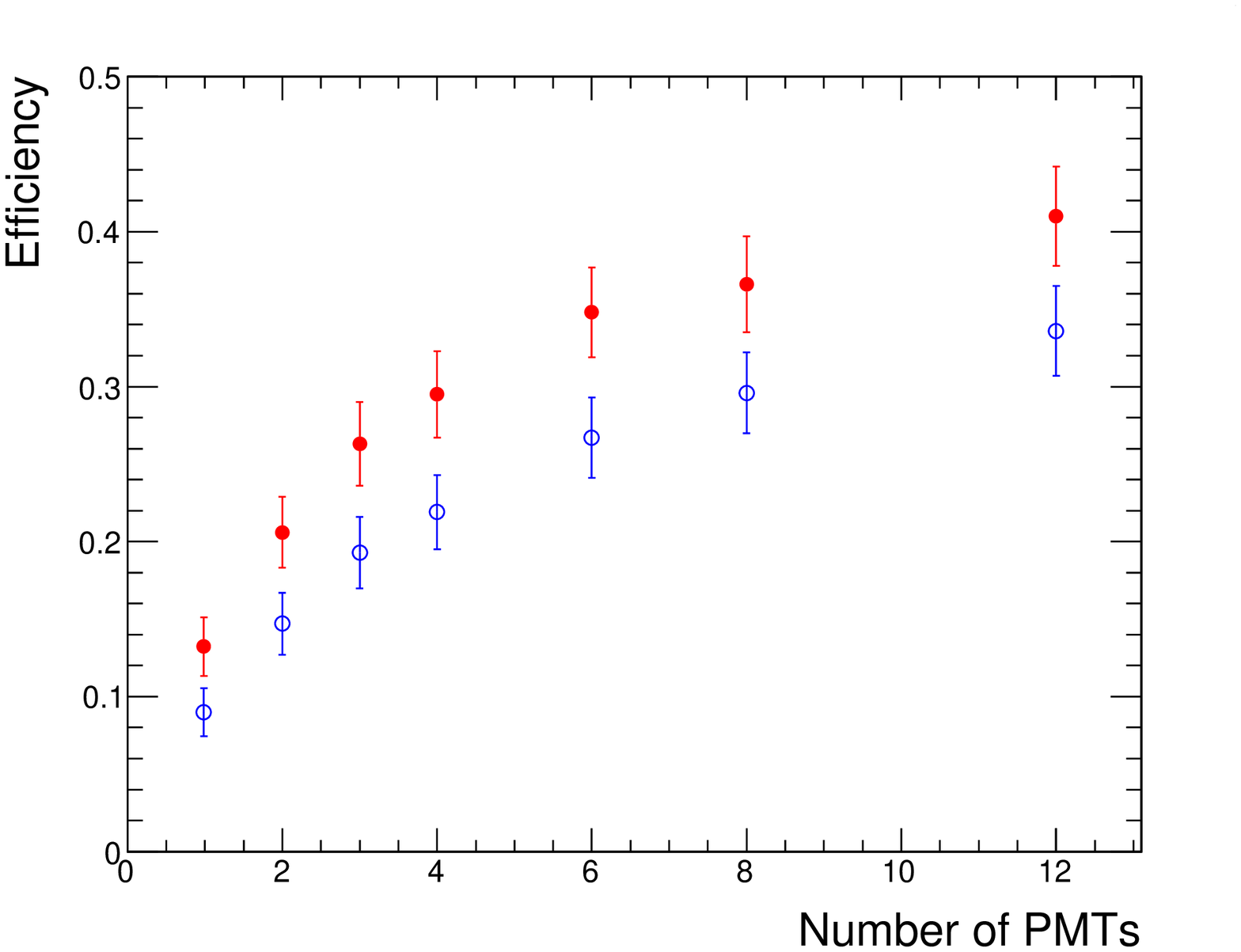}
    \includegraphics[width=0.49\textwidth]{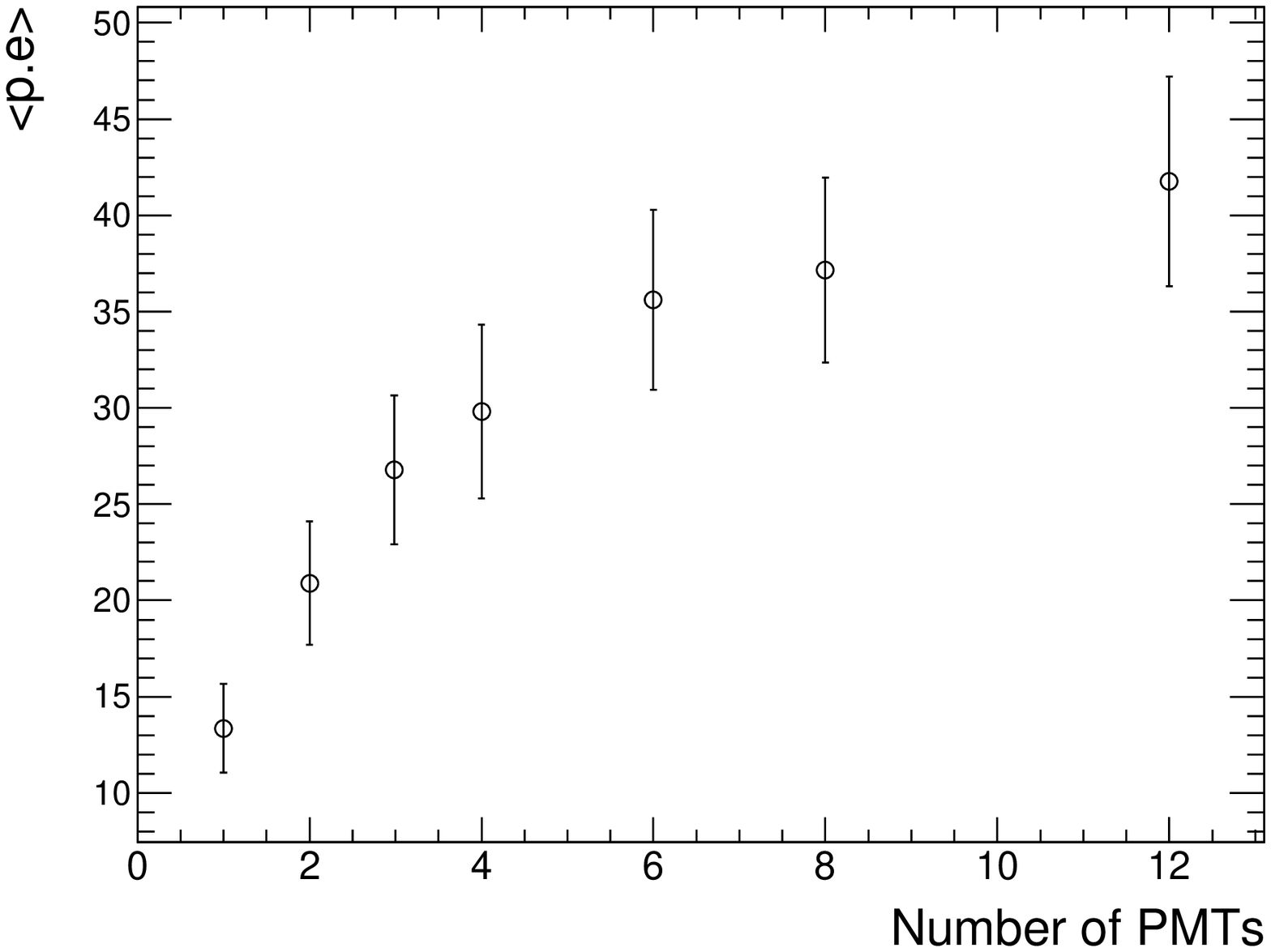}
    \caption{(Left) Photon acceptance efficiency for various PMT configurations. Open blue circles represent 8-inch-diameter PMTs, while the red solid circles are for PMTs $10$-inch-diameter. (Right) The average number of photoelectrons using the $10$-inch-diameter PMTs as a function of the number of PMT. 2 MeV positrons originating at the center of the detector have been simulated. \label{optimization}}
\end{figure}

Bottom (or top) face of the target is instrumented by six 10-inch-diameter photomultiplier tubes (PMTs) \cite{hama}. The photon acceptance efficiency of this PMT positioning configuration is found to be about $35\%$ using GEANT4 simulations \cite{GEANT4}. Figure \ref{optimization} shows photon acceptance efficiency, which is defined as the ratio between total number of photons hitting the surface of the PMTs to the total number of photons produced in target medium, for different PMT positioning configurations. Up to six PMTs are placed on bottom (or top) face of the target and PMT positioning configurations are considered like the pips on a dice. When the number of PMTs is greater than six, the PMTs are placed on both the bottom and top faces symmetrically.     

\begin{figure}[!ht]
  \centering
    \includegraphics[width=0.49\textwidth]{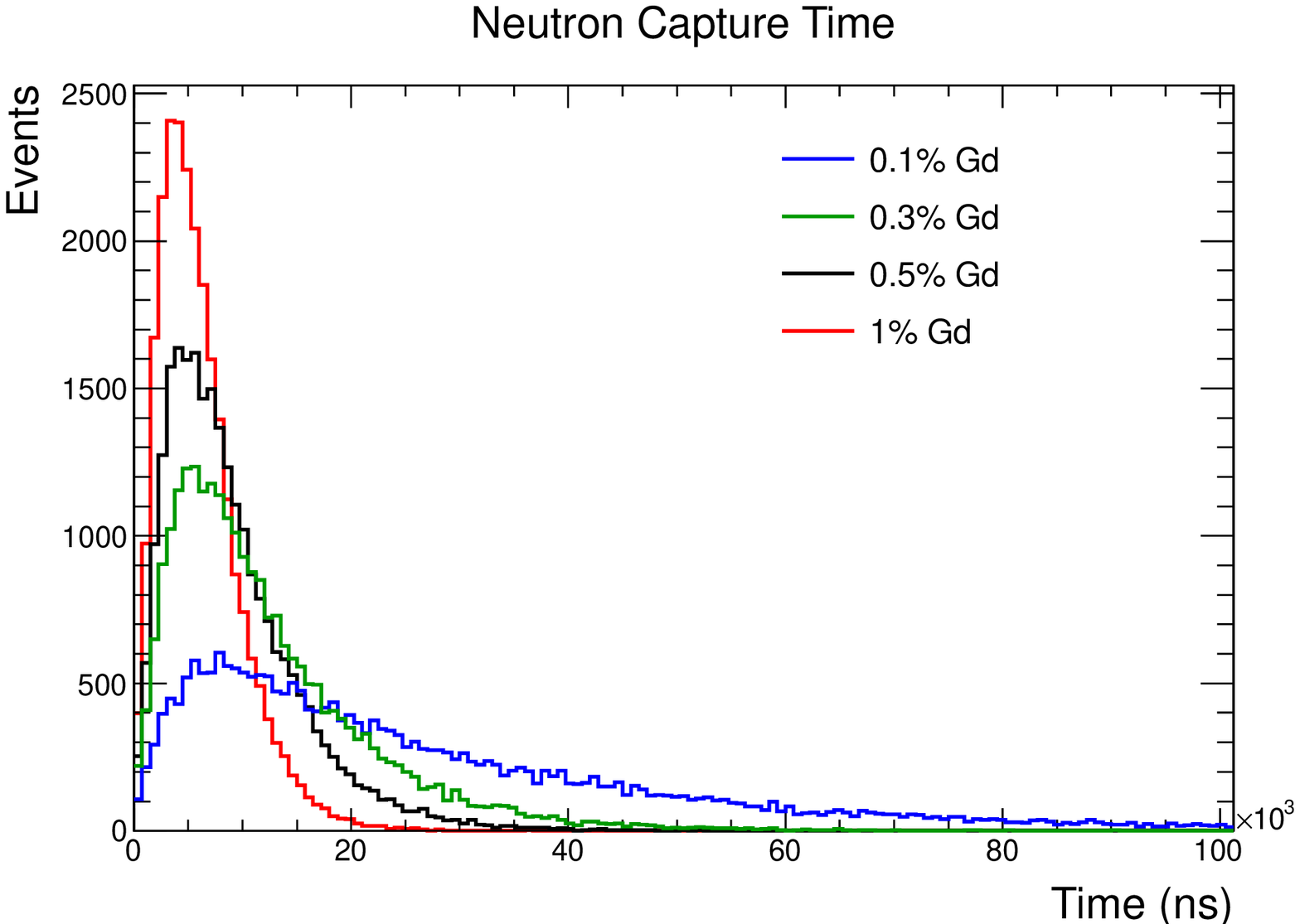}
    \includegraphics[width=0.49\textwidth]{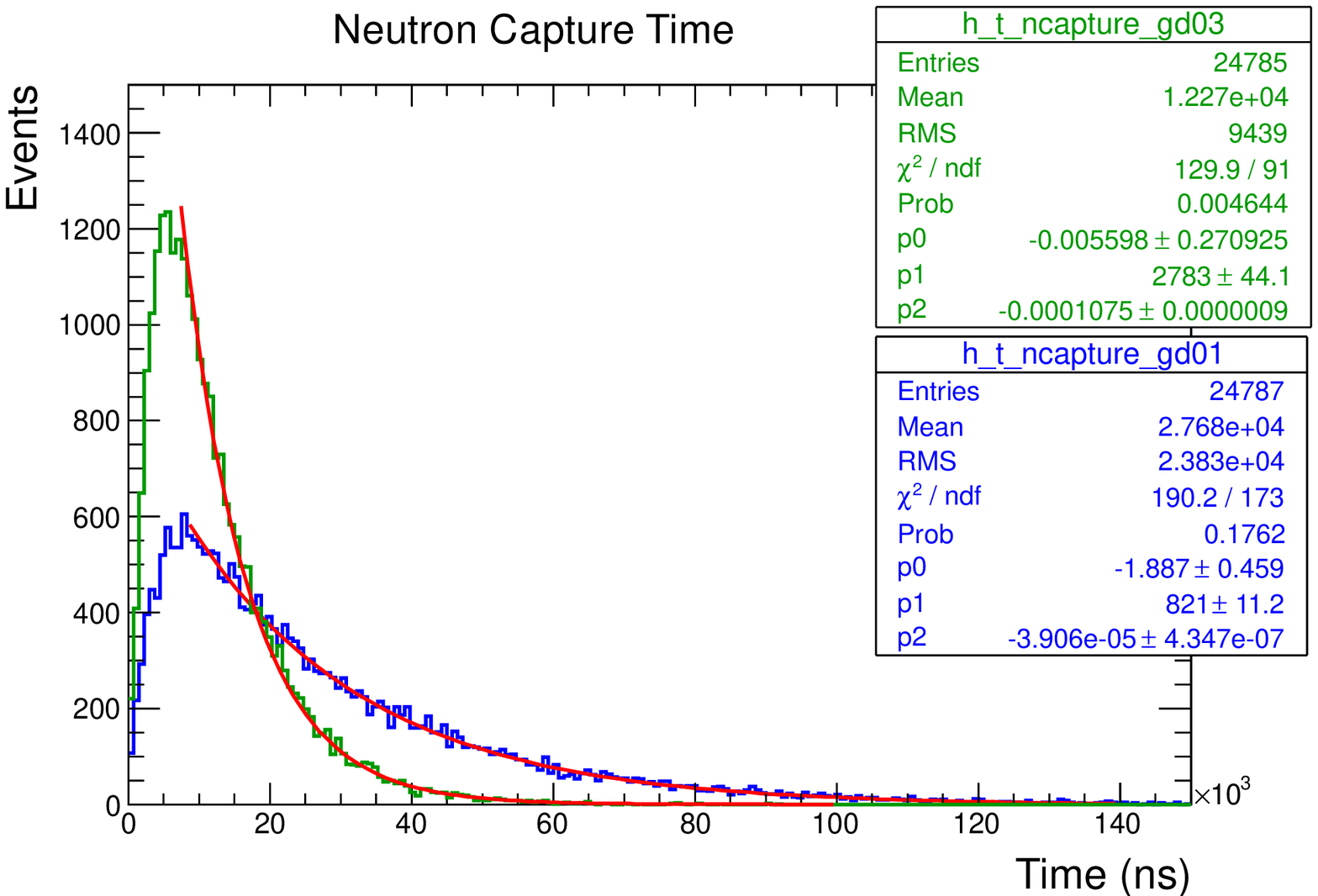}
    \caption{(Left) Time difference between prompt and delayed signals for different concentration of Gd. (Right) The distributions of $0.1\%$ and $0.3\%$ Gd concentration are fitted to an exponential function.  \label{gado}}
\end{figure}

The gadolinium (Gd) concentration in water directly effects the delayed second signal, which is caused by thermal neutron capture. This effect was studied using GEANT4 simulations. Figure \ref{gado} shows the time difference between the prompt signal and delayed signals for various Gd concentration values. The optimum Gd concentration is found to be about $0.3\%-0.5\%$. 

\begin{figure}[!ht]
  \centering
    \includegraphics[width=0.49\textwidth]{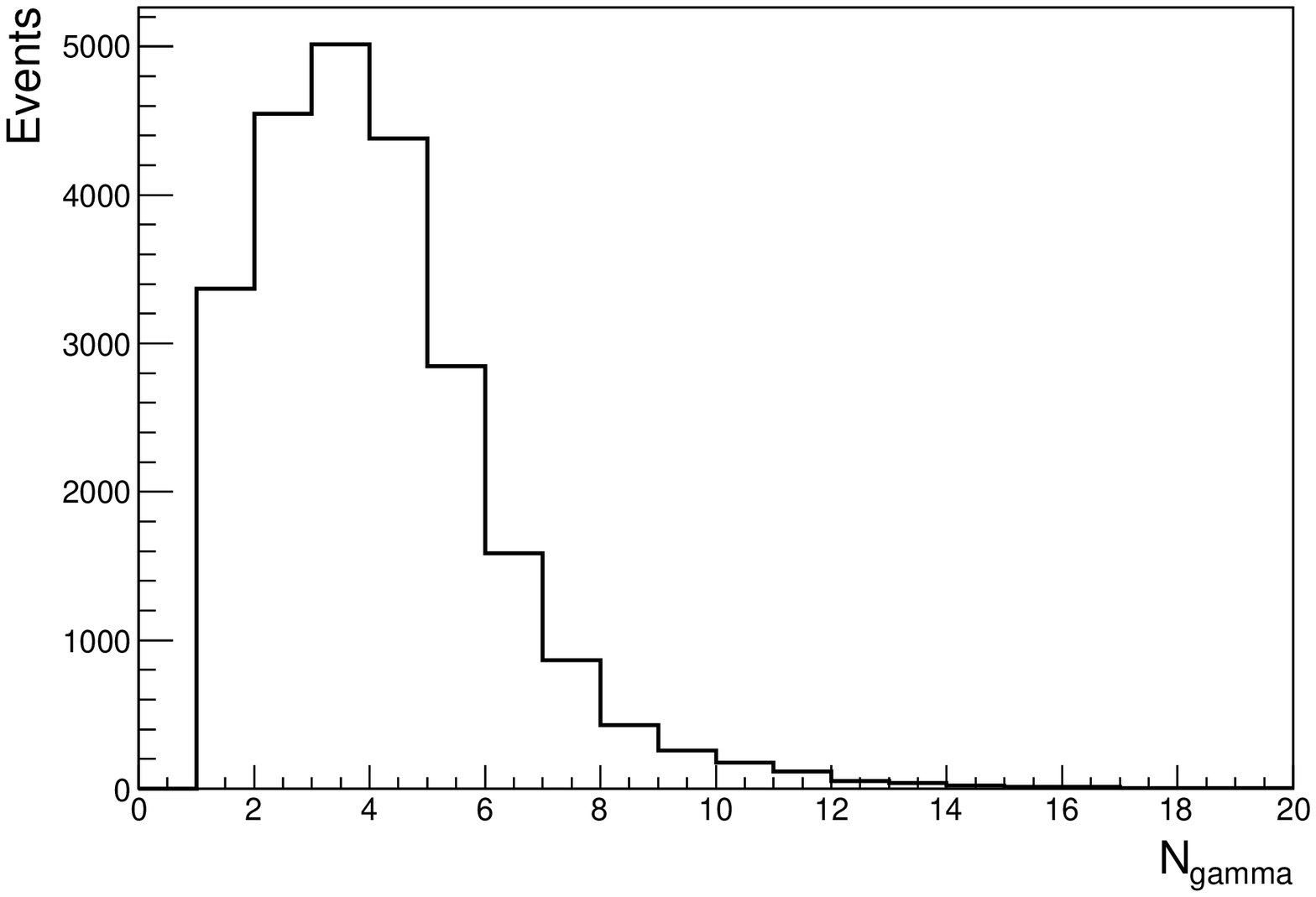}
    \includegraphics[width=0.49\textwidth]{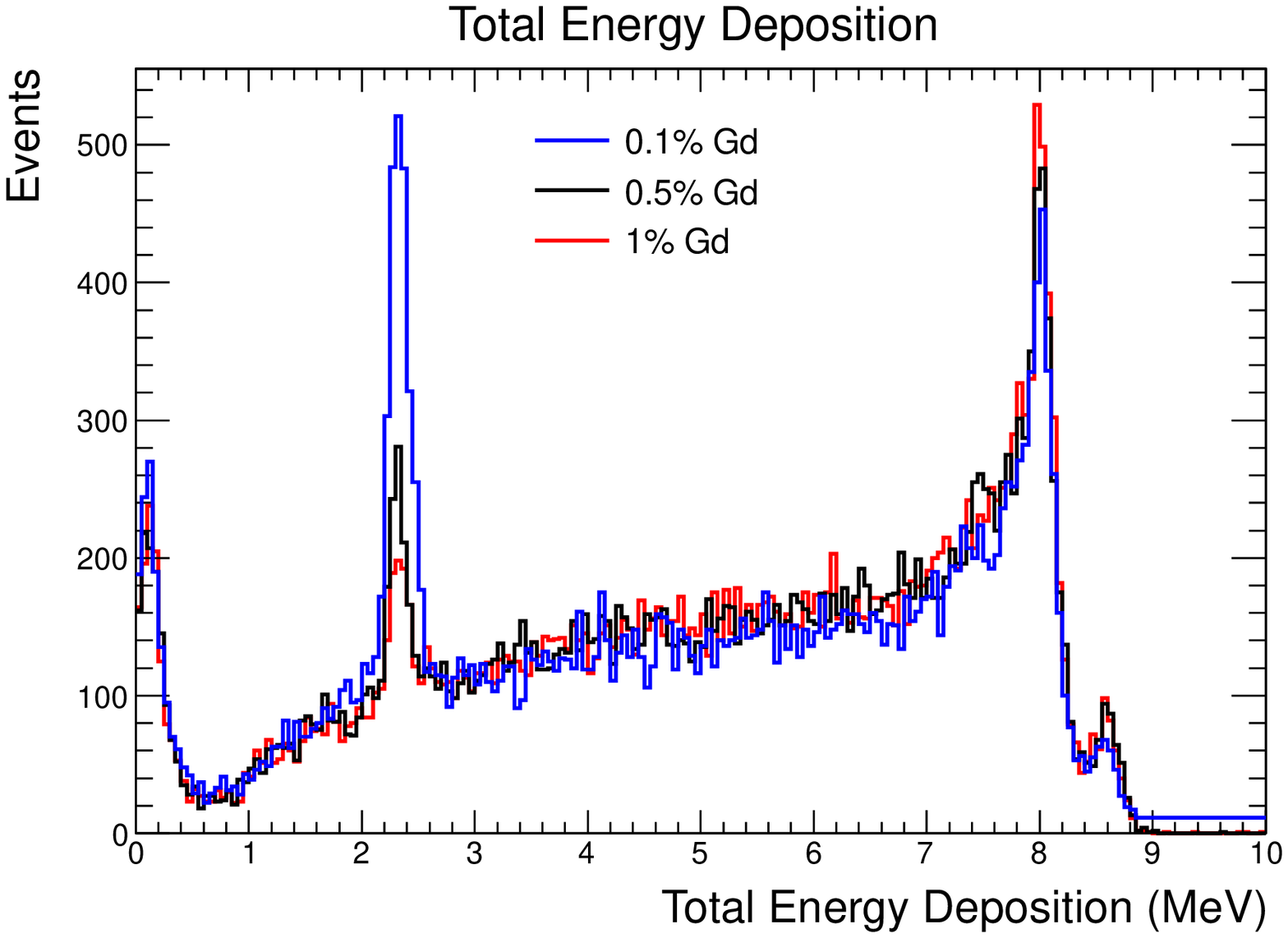}
    \caption{(Left) Total number of photons after thermal neutron capture. (Right) Total energy deposition in the medium with different Gd concentrations.  \label{notron-capture}}
\end{figure}

After thermal neutron capture by Gd, a gamma cascade is emitted with average total energy of 8 MeV. Figure \ref{notron-capture} (left) shows the total number of emitted  photons after thermal neutron capture by Gd. The right panel of the same figure presents the total deposited energy in the target medium after thermal neutron capture process. The peak around 2 MeV is produced by the thermal neutron capture from hydrogen. The peaks around 8 MeV and 8.5 MeV come from thermalization of $^{158}Gd$ and $^{156}Gd$ respectively. 

\section{Readout and Prototyping}

The reading out of a module would require a charge analog-to-digital counter (ADC) and a time-to-digital counter (TDC) with at least N channels where N is the number of photomultiplier tubes, 6 for this case. Additionally the outer section's scintillators should also be readout in order to veto events originating from cosmic rays. For a $80\times100\times120$ cm module, assuming scintillator blocks of 20 cm width and 100 and 120 cm of height, that would reqiure an additional ADC load of 26 channels. The digitizer cards, and the readout computer can all be hosted in a 6U standard VMEbus crate. The single board computer would not need to run a realtime operating system, thus a Linux based solution would be sufficient. The trigger logic can be initially setup using NIM modules, and later on can be upgraded to a faster timing and smaller footprint by using an FPGA. The data out of the VMEbus crate can be shipped off to another computer with a long term storage device via a simple gigabit switched network.
A simple detector unit could be produced and tested with minimum effort in less than two years. The calibration of the detector can be easily achieved using a low energy electron or proton beam.
A project is being submitted to TUBITAK for funding to produce a demonstration module.

\section{Conclusions}
In this paper, a first study for the design of a monitoring detector facility for Akkuyu NPP has been outlined. Simulations with GEANT4 have identified that 6 PMTs of 10inch diameter active area would gather about 35\% of photons produced in the detector. The same simulation study has shown that the active volume of the detector itself could be made from $0.3\%-0.5\%$ Gd-doped water. The final design, construction and commissioning of a monitoring unit is expected to take up to two years. Many such units could be combined together to increase the event yield and thus the monitoring efficiency.


\end{document}